\documentclass[showpacs,preprintnumbers,amsmath,amssymb,aps,prb,floatfix,reprint,superscriptaddress]{revtex4-1}

\usepackage{graphicx}
\usepackage{hyperref}

\DeclareMathOperator{\gd}{gd} 
\renewcommand{\vec}[1]{{\bf #1}}

\begin{document}

\title{The effect of boundary-induced chirality on magnetic textures in thin films}
\author{Jeroen Mulkers}
\email[Email: ]{jeroen.mulkers@uantwerpen.be}
\affiliation{Department of Physics, University of Antwerp, Antwerp, Belgium}
\affiliation{DyNaMat, Department of Solid State Sciences, Ghent University, Ghent, Belgium}
\author{Kjetil M. D. Hals}
\affiliation{Institute of Physics, Johannes Gutenberg Universit\"at, 55128 Mainz, Germany}
\affiliation{Department of Engineering and Science, University of Agder, 4879 Grimstad, Norway}
\author{Jonathan Leliaert}
\affiliation{DyNaMat, Department of Solid State Sciences, Ghent University, Ghent, Belgium}
\author{Milorad V. Milo\v{s}evi\'c}
\email[Email: ]{milorad.milosevic@uantwerpen.be}
\affiliation{Department of Physics, Antwerp University, Antwerp, Belgium}
\author{Bartel Van Waeyenberge}
\affiliation{DyNaMat, Department of Solid State Sciences, Ghent University, Ghent, Belgium}
\author{Karin Everschor-Sitte}
\affiliation{Institute of Physics, Johannes Gutenberg Universit\"at, 55128 Mainz, Germany}

\begin{abstract}
In the quest for miniaturizing magnetic devices, the effects of boundaries and surfaces become increasingly important. Here we show how the recently predicted boundary-induced Dzyaloshinskii-Moriya interaction (DMI) affects the magnetization of ferromagnetic films with a $C_{\infty v}$ symmetry and a perpendicular magnetic anisotropy. For an otherwise uniformly magnetized film, we find a surface twist when the magnetization in the bulk is canted by an in-plane external field. This twist at the surfaces caused by the boundary-induced DMI differs from the common canting caused by internal DMI observed at the edges of a chiral magnet. Further, we find that the surface twist due to the boundary-induced DMI strongly affects the width of the domain wall at the surfaces. We also find that the skyrmion radius increases in the depth of the film, with the average size of the skyrmion increasing with boundary-induced DMI. This increase suggests that the boundary-induced DMI contributes to the stability of the skyrmion.
\end{abstract}


\maketitle

\section{Introduction}

That boundary conditions (BCs) have important consequences pervades many areas of physics. Well-known examples are the discrete frequencies of a vibrating string clamped at the edges or different electrostatic solutions that arise in either Dirichlet BCs or Neumann BCs. The effects of boundaries are also crucial in micromagnetism and have been addressed already more than 20 years ago, e.g.\ see Ref.~\onlinecite{Labrune1995}. In more recent years, magnetic systems with broken space inversion symmetry have attracted a lot of attention. The broken inversion symmetry allows for the Dzyaloshinskii-Moriya interaction (DMI) which twists magnetic textures in a chiral way. Therefore, such systems can host novel topological magnetic textures like chiral domain walls\cite{Klaui2005,Parkin2008,Miron2011,Ryu2013,Emori2013} and magnetic skyrmions\cite{Pokrovsky1979,Bogdanov1989,Muhlbauer2009,Bogdanov1994}. In these systems, it has been shown that the DMI induces canting of the magnetization at the edge \cite{Rohart2013,Meynell2014}. This canting can have a profound effect on the confinement of modulated magnetic textures like helices and skyrmions \cite{Wilson2013,Leonov2016,Muller2016,Mulkers2016,Leonov2017}.

\begin{figure}
    \includegraphics{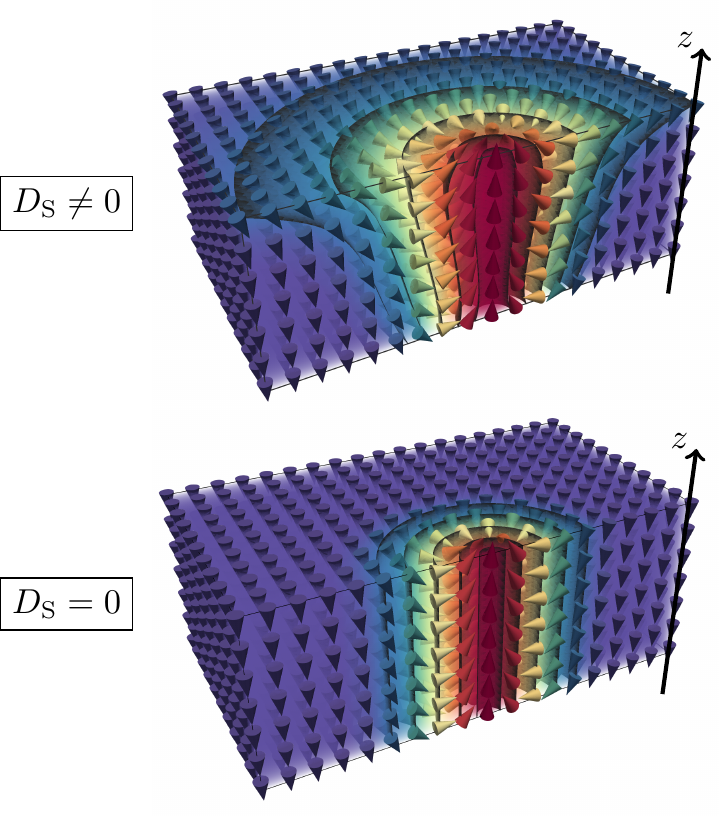} 
    \caption{Cross section of a skyrmion in an extended film with boundary-induced DMI ($D_{\text{S}}\neq0$) at the top and bottom surface, and without boundary-induced DMI ($D_{\text{S}}=0$) . Both systems are shown on the same scale. The dark contours represent isomagnetizations.}
    \label{fig:demo}
\end{figure}

In this manuscript, we exploit the effect of a recently discovered contribution to the boundary condition in systems with generalized DMI\cite{Hals2017}. It has been shown in Ref.~\onlinecite{Hals2017}, by means of symmetry analysis, that this boundary condition with generalized DMI requires the full tensorial structure of the third-rank DMI tensor and not just the antisymmetric part. Already for the case of a simple ferromagnetic thin film, a new type of DMI-induced spin structure, i.e.\ a purely boundary-driven magnetic twist state at the top and bottom surface, was predicted analytically. In this work, we study numerically and analytically the effect of this type of boundary-induced DMI on the ferromagnetic state, on a domain wall, and on a magnetic skyrmion in a ferromagnetic film with $C_{\infty v}$ symmetry. The paper is structured as follows. In Sec.~\ref{sec:model}, we describe the system and review the idea of the boundary-induced DMI. In Sec.~\ref{sec:uniform}, we consider the effect of such a term in an otherwise uniform state. We do find a chiral edge canting and analyze it thoroughly in Sec.~\ref{sec:edge}. Subsequently, we study the effect of the boundary-induced DMI on isolated domain walls and magnetic skyrmions in Sec.~\ref{sec:wall} and \ref{sec:skyrmion}, respectively. We find that boundary induced DMI leads to an increase/decrease of the domain wall width at the top/bottom surface of the sample and a depth-dependent increase of the skyrmion radius, as shown in Fig.~\ref{fig:demo}. Our results are summarized in Sec.~\ref{sec:conclusions}.

\section{Model and Methods\label{sec:model}}

We model a thin magnetic film (in the $xy$-plane) with perpendicular magnetic anisotropy (PMA) and generalized DMI. The free energy functional of the magnetization $\vec{M}(\vec{r}) = M_{\text{s}}\vec{m}(\vec{r})$ with a constant saturation magnetization~$M_{\text{s}}$ is given by $E= \int \varepsilon \,dV$ for which the energy density is
\begin{equation}
\begin{split}
    \varepsilon = A(\nabla \vec{m})^2 - K m_z^2
    +\sum_{\mu\alpha\beta} D^{\mu}_{\alpha\beta} m_{\alpha} \partial_{\mu} m_{\beta} \\
    - \mu_0 \vec{M}\cdot \vec{H}^{\text{ext}}
    - \textstyle{\frac{1}{2}}\mu_0 \vec{M} \cdot \vec{H}^{\text{demag}},
    \label{eq:energydensity}
\end{split}
\end{equation}
with exchange stiffness $A$, applied field $\vec{H}^{\text{ext}}$, anisotropy constant $K$, and $3^3=27$ DMI tensor elements $D^{\mu}_{\alpha\beta}$. The indices $\alpha$ and $\beta$ denote the components of the magnetization direction, whereas the index $\mu$ denotes the component of the magnetization gradient.

The most prominent effect of the demagnetization can be translated into a shape anisotropy. For a laterally infinite film with a magnetization $\vec{M}(z)$, which varies only along the depth of the film, one can prove that the use of the effective anisotropy describes the demagnetization exactly [see Appendix~\ref{sec:demaguniform}]. Therefore, we do not calculate the demagnetization energy explicitly. Instead, we approximate the demagnetization by lowering the anisotropy parameter~$K$ to the effective anisotropy parameter $K_e = K-\mu_0 M_{\text{s}}^2/2$. For the chiral states, such as domain walls and skyrmions, we expect this approximation to be valid for thin films, but it becomes suboptimal for very thick films. In this case, full-blown 3D micromagnetic simulations, which include full computation of the dipolar interactions, are needed.

We study films with a thickness~$d$ larger than the typical length scale~$\xi=\sqrt{A/K_{\text{e}}}$. One could expect that the effect of the boundary-induced DMI in very thin films $(d \ll \xi)$ is even more pronounced. However, we do not consider such thin films in this paper, because the micromagnetic framework is not suited to study a varying magnetization on dimensions smaller than the typical length scale~$\xi$. For these thin films, one should instead resort to an atomistic description of the film, which lies outside the scope of the present analysis.

To reduce the number of free parameters, we can eliminate the exchange stiffness $A$ and the effective anisotropy constant $K_{\text{e}}$ by expressing relevant quantities in units of the length scale $\xi$ and the critical DMI strength $D_c=4\sqrt{AK_{\text{e}}}/\pi$. Furthermore, we will express magnetic fields strengths in units of the critical field $h_c = 2K_{\text{e}}/\mu_0 M_{\text{s}}$. Using the demagnetization approximation and the above definitions, we obtain the following rescaled energy density:
\begin{equation}
\begin{split}
    \frac{\varepsilon}{K_{\text{e}}} = (\xi\nabla \vec{m})^2 - m_z^2
    - 2 \vec{m}\cdot \frac{\vec{H}^{\text{ext}}}{h_c} \\
    +\sum_{\mu\alpha\beta}\frac{4}{\pi} \frac{D^{\mu}_{\alpha\beta}}{D_{\text{c}}} m_{\alpha} \xi\partial_{\mu} m_{\beta}.
\end{split}
\end{equation}

In this paper, we concentrate our discussion on magnetic thin films with $C_{\infty v}$ symmetry. In this case, there are only four independent tensor coefficients in the DMI tensor: $D^x_{xz}=D^y_{yz}$, $D^z_{xx} = D^{z}_{yy}$, $D^{x}_{zx} = D^y_{zy}$, and $D^z_{zz}$. The remaining tensor coefficients vanish because of the symmetry\cite{Hals2017}. The DMI energy density terms given in Eq.~\eqref{eq:energydensity} can be rearranged in terms which are symmetric and antisymmetric for a permutation of the magnetization components ($\alpha\leftrightarrow\beta$). According to this rearrangement, we redefine the four independent tensor coefficients by three symmetric DMI parameters~$D^{(1)}_{\text{S}}$, $D^{(2)}_{\text{S}}$, $D^{(3)}_{\text{S}}$, and an asymmetric DMI parameter~$D_{\text{A}}$, as follows:
\begin{subequations}
\begin{align}
    D^x_{xz} = D^y_{yz} &\equiv D^{(1)}_{\text{S}} + D_{\text{A}}, \label{eq:dmiparams1} \\
    D^x_{zx} = D^y_{zy} &\equiv D^{(1)}_{\text{S}} - D_{\text{A}}, \label{eq:dmiparams2} \\
    D^z_{xx} = D^{z}_{yy} &\equiv D^{(2)}_{\text{S}}, \label{eq:dmiparams3} \\
    D^z_{zz} &\equiv D^{(3)}_{\text{S}}.\label{eq:dmiparams4}
\end{align}
\end{subequations}
Because the magnetization~$m$ is normalized, the symmetric DMI parameters~$D^{(2)}_{\text{S}}$ and $D^{(3)}_{\text{S}}$ can be combined using
\begin{equation}
	D_{\text{S}} \equiv D^{(2)}_{\text{S}} - D^{(3)}_{\text{S}}
\end{equation}
in the free energy, without losing generality. Thus, only three relevant DMI parameters remain: $D_{\text{A}}$, $D_{\text{S}}$, and $D^{(1)}_{\text{S}}$.

We will refer to the antisymmetric DMI terms as the \emph{internal DMI} because these DMI terms contribute to the energy density within the bulk of the film. This internal DMI can be found either in the bulk of a ferromagnetic lattice without space inversion symmetry, or it can be induced in an ultra-thin ferromagnetic layer coupled to a nonmagnetic layer with a strong spin-orbit coupling\cite{Crepieux1998} (commonly called the interfacially-induced DMI). In contrast, the symmetric DMI terms lead to an energy contribution which depends only on the surface magnetization. To demonstrate this, we integrate the energy density terms related to $D^{\mu}_{\alpha\beta}$ and $D^{\mu}_{\beta\alpha}$ over $\mu$:
\begin{eqnarray}
    &&\int^c_a [ D^{\mu}_{\alpha\beta} m_{\alpha} \partial_{\mu} m_{\beta} + D^{\mu}_{\beta\alpha} m_{\beta} \partial_{\mu} m_{\alpha} ] \,d \mu \nonumber \\
    &&\begin{split}
    \quad\quad\quad = \frac{D^{\mu}_{\alpha\beta} - D^{\mu}_{\beta\alpha}}{2}\int_a^c [ m_{\alpha} \partial_{\mu} m_{\beta} - m_{\beta} \partial_{\mu} m_{\alpha} ]\,d\mu  \\
     \quad\quad +  \frac{D^{\mu}_{\alpha\beta}+ D^{\mu}_{\beta\alpha}}{2} \left. m_{\alpha} m_{\beta} \right|_{\mu=a}^{\mu=c}.
    \end{split}
\end{eqnarray}
Note that the symmetric part depends only on the magnetization at the surfaces $\mu=a$ and $\mu=c$. Therefore, we will refer to the symmetric DMI terms as the \emph{boundary-induced DMI}. In our system, this means that the energy term related to $D_{\text{S}}$ depends only on the magnetization at the top~$t$ and bottom~$b$ surface of the film. The energy term related to $D^{(1)}_{\text{S}}$ depends only on the magnetization at the edges (lateral surfaces) of a finite film and is irrelevant in case of a laterally infinite film.

In the following, we will scrutinize the effect of the boundary-induced DMI at the top and the bottom surface on selected magnetic textures including chiral domain walls and skyrmions. As a general procedure for all examples studied, we will first simplify the energy density of Eq.~\eqref{eq:energydensity} by taking into account the symmetry of the specific system. If possible, we minimize the energy functional analytically using variational calculus. Complementary, we resort to a numerical minimization of the energy. To this end, we discretize the simplified expression for the free energy and calculate its gradient with respect to the magnetization at each grid point. The used grid size is $0.02\xi$ in the $z$ direction and $0.1\xi$ in the $xy$-plane unless otherwise mentioned. Using the Barzilai-Borwein gradient method, we minimize the energy starting from an initial guess of the equilibrium state\cite{Barzilai1988}.

\section{Quasi-uniform state\label{sec:uniform}}

In this section, we study the surface twist of an otherwise uniformly magnetized film, which is described by the model of Eq.~\eqref{eq:energydensity} with an external in-plane magnetic field $\vec{H}^{\text{ext}}$. The anisotropy constant $K_{\rm e}$ is assumed to be positive, $K_{\text{e}} > 0$, so that the magnetization is collinear with the $z$-axis in absence of $\vec{H}^{\text{ext}}$. As we consider a laterally infinite film, the equilibrium state close to the uniform state will not vary along the $x$ or $y$~direction, but a surface twist in the $z$~direction is to be expected due to the boundary-induced DMI\cite{Hals2017}. Under these considerations, the use of an effective anisotropy captures the effects of the demagnetization exactly [see Appendix~\ref{sec:demaguniform}]. We rotate the coordinate system so that the $x$~axis is aligned with the applied field, i.e.\ $\vec{H}^{\text{ext}} = h\hat{e}_x$ with $h>0$. Now, the in-plane field cants the magnetization in the $x$~direction away from the film normal ($z$~direction). Hence, the equilibrium magnetization is fully described by the angle~$\psi(z)$, which describes the tilting away from the $x$~axis:
\begin{equation}
    \vec{m}(z)=(\cos\psi(z),0,\sin\psi(z)).
    \label{eq:m}
\end{equation}

The above model contains only three relevant free parameters: the applied field~$h$, the symmetric DMI parameter~$D_{\text{S}}$, and the film thickness~$d$. Note that the value of the antisymmetric DMI parameter~$D_{\text{A}}$ is irrelevant in this case since the magnetization does not vary in the $x$ and $y$~direction [see upper indices in Eqs.~\eqref{eq:dmiparams1} and \eqref{eq:dmiparams2}]. Furthermore, we will assume that $D_\text{S}$ is positive without losing generality: when switching the sign of $D_{\text{S}}$, the role of the top and bottom surface simply interchange. Combining the energy density with the magnetization ansatz of Eq.~\eqref{eq:m} leads to
\begin{equation}
\begin{split}
    \varepsilon = A( \partial_z \psi)^2 - 2 D_{\text{S}} \sin\psi\cos\psi\ \partial_z\psi   \\
    - K_{\text{e}} \sin^2\psi - M_{\text{s}} h\cos\psi \label{eq:energy1D}.
\end{split}
\end{equation}
Minimizing the free energy per surface area~$\int \varepsilon\,d z$ yields an expression for the magnetization profile. We first calculate the case without boundary-induced DMI ($D_\text{S}=0$) being identical to the bulk value $\psi_{\text{B}}$ of a thick film with boundary-induced DMI. For this case, we obtain
\begin{equation}
    \cos\psi_{\text{B}} = \frac{h}{h_{\text{c}}}
    \ \ \ \text{for}\ \ h < h_{\text{c}} = \frac{2K_{\text{e}}}{\mu_0 M_{\text{s}}}.
    \label{eq:psibulk}
\end{equation}
For an in-plane field with strength $h$ larger than the critical field strength~$h_{\text{c}}$ the magnetization is fully in plane, i.e.\ $\psi_B=0$. In the absence of an external in-plane field, the magnetization in the bulk is oriented out of plane, either up or down, so $\psi_{\text{B}}=\pm \pi/2$.

Let us now consider the case with a non-vanishing boundary-induced DMI. As we will show below, an external in-plane field and a nonzero symmetric DMI strength~$D_{\text{S}}$ are the prerequisites for the occurrence of surface twists in the quasi-uniform state in films. The minimization of the energy leads to the following second order differential equation for~$\psi(z)$:
\begin{equation}
    \xi^2\partial^2_z \psi + \cos\psi\sin\psi - \frac{h}{h_{\text{c}}}\sin\psi = 0 \label{eq:DEminuniform},
\end{equation}
in combination with the Neumann BC\cite{Hals2017}
\begin{equation}
    \left.\xi\partial_z\psi\right|_{t,b} = \frac{2D_{\text{S}}}{\pi D_c}\sin(2\psi_{t,b}).
    \label{eq:BC}
\end{equation}
In what follows, we will study the thick film limit analytically and resort to a numerical minimization of the free energy for films with a finite thickness.

\subsection{Thick film limit ($d\rightarrow\infty$)\label{sec:uniformthick}}

To study this case, we consider a ferromagnet which takes up half the space. We fill the upper half of the space ($z>0$) to study the bottom surface of the thick film, and the bottom half ($z<0$) for the respective consideration as the top surface. We need to handle both cases explicitly because the surface DMI term removes the symmetry between the magnetization profiles at the top and bottom surface. To exploit the effects of the BC, we integrate Eq.~\eqref{eq:DEminuniform} over $\psi$, starting from the magnetization in the bulk $\psi(z=\pm\infty) = \psi_{\text{B}}$ and obtain an equation that is valid for a surface at the $z=0$ surface in both cases:
\begin{equation}
    (\xi\partial_z\psi)^2 = \cos^2\psi - \cos^2\psi_{\text{B}} -2\frac{h}{h_{\text{c}}} (\cos\psi-\cos\psi_{\text{B}}).\label{eq:thickDE}
\end{equation}
Here, we assumed that the magnetization is uniform in the bulk, and thus $\left.\partial_z\psi\right|_{\pm \infty} = 0$. Filling in the BC [Eq.~\eqref{eq:BC}] in the l.h.s.\ of Eq.~\eqref{eq:thickDE} yields an equation for the surface magnetization angle at~$\psi(z=0)$, which has four solutions~$\pm\psi_{t,b}$ in total. Here, we use the bottom indices $t$ and $b$ to indicate the solution at the top and bottom surface respectively. The positive values are the solutions for $\psi_{\text{B}}\geq0$ and the negative solutions for $\psi_{\text{B}}\leq0$. Assuming $\psi(z)$ is monotonic, we can derive the inequalities \begin{equation}
    0 \leq |\psi_b| \leq |\psi_{\text{B}}| \leq |\psi_t| \leq \frac{\pi}{2}
\end{equation}
from the BC, which can then be used to distinguish the different solutions.

\begin{figure}
    \centering
    \includegraphics{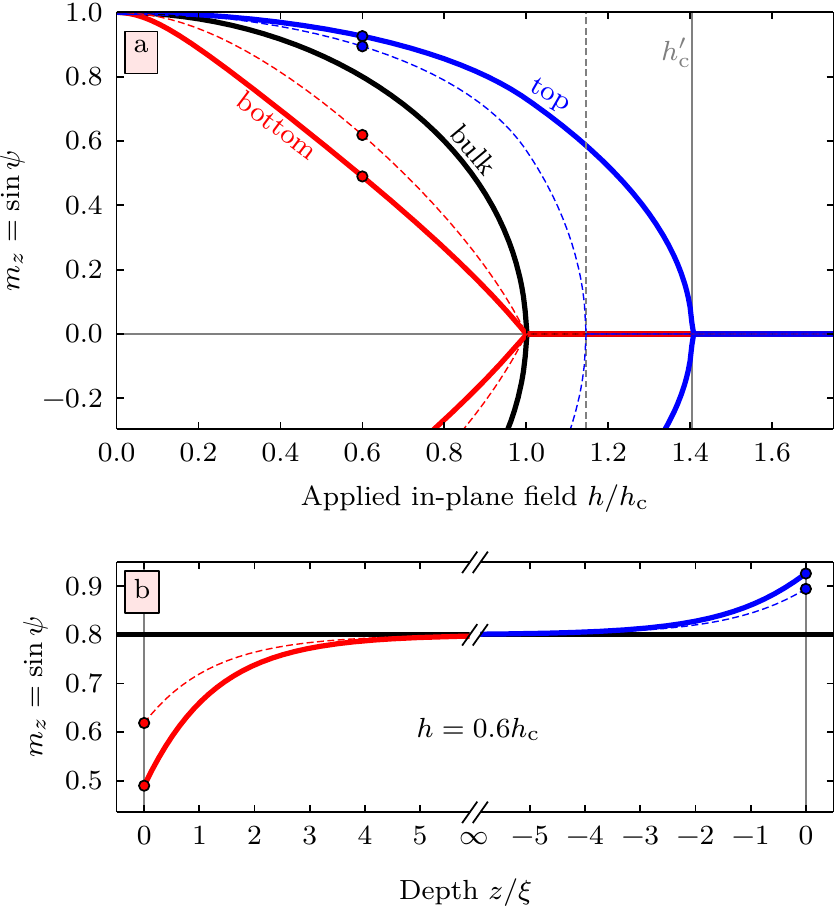} 
    \caption{(a) Out-of-plane magnetization~$m_z$ at the top surface (blue), the bottom surface (red), and in the bulk (black) of a very thick film as a function of an external in-plane field~$h\hat{e}_x$ for symmetric DMI parameter $D_{\text{S}}=0.5D_c$ (solid line) and $D_{\text{S}}=0.2D_c$ (dashed line). (b) The magnetization profile~$m_z(z)$ near the top and near the bottom surface of a thick film for an external in-plane field $h=0.6h_{\text{c}}$. The surface is positioned at $z=0$ in both cases. The same coloring and line style is used as in (a). The dots in two panels denote the same top and bottom surface magnetization.}
    \label{fig:uniforminfinite}
\end{figure}

Fig.~\ref{fig:uniforminfinite}(a) shows the out-of-plane magnetization at the surfaces~$m_z=\pm\sin\psi_{t,b}$ as a function of the applied field. Note that the magnitudes of the surface twist at the top and bottom surface are in general different from each other: $|\psi_t - \psi_{\text{B}}| \neq |\psi_b - \psi_{\text{B}}|$. In Ref.~\onlinecite{Hals2017}, Hals \emph{et al.} studied an equivalent problem to first order in the DMI parameter $D_S$. In this limiting case, they found a symmetric estimation of the surface twist at the top and bottom surface.

In this model, there are two critical field strengths. First there is the critical field strength~$h_{\text{c}}$ already defined in Eq.~\eqref{eq:psibulk}, which turns the bulk magnetization fully in plane. One can deduce from the BC in Eq.~\eqref{eq:BC} that if the bulk magnetization is in plane then also the magnetization at the bottom surface is fully in plane. The second critical field strength~$h'_{\text{c}}$ can be defined as the minimal field strength needed to turn the magnetization at the top surface fully in plane. This corresponds to a solution that has an in-plane magnetization everywhere in the film, i.e.\ $\psi(z)=0$. Filling in the BC and an infinitely small magnetization angle in Eq.~\eqref{eq:thickDE} yields an expression for the second critical field:
\begin{equation}
    h'_{\text{c}} = \min_{\psi(z)=\pi/2} h = h_{\text{c}}\left(1+ \left(\frac{4 D_{\text{S}}}{\pi D_{\text{c}}}\right)^2\right).
\end{equation}
The increase of the critical field strength~$h'_{\text{c}}$ with the boundary-induced DMI strength~$D_{\text{S}}$ can be intuitively understood by noting that a stronger boundary-induced DMI leads to a larger surface twist, and consequently, a larger in-plane field is needed to turn the top surface magnetization fully in plane. For in-plane field strengths between the two critical values, the magnetization in the bulk and at the bottom surface is in-plane while the magnetization at the top surface has an out-of-plane component. In this case, the surface twist at the top is degenerate in the sense that the out-of-plane component can either be positive or negative while having the same minimal energy.

Integration of Eq.~\eqref{eq:thickDE}, starting from the magnetization at the open surface, yields an implicit equation for the magnetization profile $\psi(z)$:
\begin{equation}
    z = \int^{\psi(z)}_{\pm\psi_{t,b}} \frac{d\psi'}{\sqrt{
    \cos^2\psi'-\cos^2\psi_{\text{B}}-2\frac{h}{h_{\text{c}}}(\cos\psi'-\cos\psi_{\text{B}})}}.
\end{equation}
The profile of the surface twist near the top and bottom surface is shown in Fig.~\ref{fig:uniforminfinite}(b). The gradient is positive at the bottom and the top surface, as is expected for a positive boundary-induced DMI.  The figure also shows the asymmetry in the solutions for the bottom and the top surface.

\subsection{Finite thickness}

\begin{figure}
    \centering
    \includegraphics{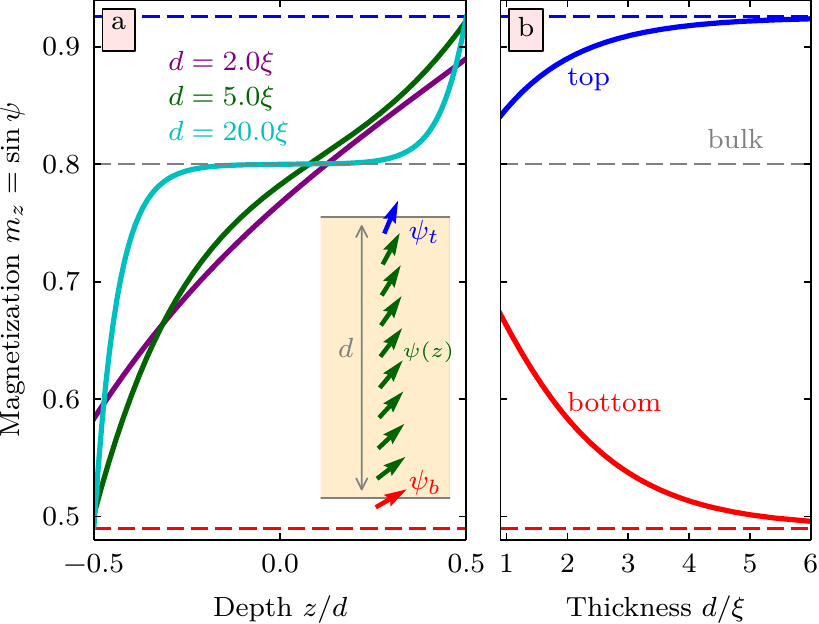} 
    \caption{(a) The magnetization profile $m_z(z) = \sin\psi(z)$ in films of different thickness $d$, for symmetric DMI parameter $D_{\text{S}}=0.5D_c$, and in-plane field $h=0.6 h_{\text{c}}$, obtained after a numerical minimization of the free energy. The inset shows a sketch of the magnetization exhibiting surface twists in a film with finite thickness $d$. (b) The magnetization at the top(bottom) surface as a function of the film thickness $d$. Dashed lines represent the asymptotic values for $d\rightarrow \infty$, which were calculated analytically.}
    \label{fig:uniformfinite}
\end{figure}

Calculating the magnetization profile of the quasi-uniform state in a film with a finite thickness $d>0$ is challenging to approach analytically. That is why we resort to a numerical minimization of the free energy given in Eq.~\eqref{eq:energy1D}. The obtained magnetization angle profiles~$\psi(z)$ are shown in Fig.~\ref{fig:uniformfinite}(b) for films of different thicknesses~$d$. Also here, we see the asymmetry in the twist at the top and bottom surface of thick films. Figure~\ref{fig:uniformfinite}(c) shows the twists at the top surface~$\psi_t$ and the bottom surface~$\psi_b$ as a function of the thickness~$d$. For thick films, the twists at the surfaces converges to the analytical result of the thick film limit, derived in Sec.~\ref{sec:uniformthick}

\section{Edge canting\label{sec:edge}}

From the Neumann BC given in Eq.~\eqref{eq:BC}, one can deduce that a surface twist in the ferromagnetic ground state of a laterally infinite film only occurs when the magnetization is canted away from the easy axis. In the previous section, we showed that such canting can be achieved by applying an in-plane field. In films with limited lateral size, this canting away from the easy axis occurs at the edges without applied field due to the internal DMI \cite{Rohart2013,Meynell2014}. Consequently, one could expect that even without an applied field, surface twists also can occur close to the sample edges.

When we minimize the energy analytically without boundary-induced DMI, we find that the canting of the magnetization along the $x$~direction close to the right edge $(x=a)$ of a finite size film is given by
\begin{equation}
    \psi(x) = 2 \arctan \left[ \mathrm{e}^{-(x-a)/\xi} \tan\left( \frac{\psi_0}{2} + \frac{\pi}{4} \right) \right] - \frac{\pi}{2}
\end{equation}
for $x<a$. Here, $\psi_0 = \arccos(2 D_{\text{A}}/\pi D_c) $ is the canting angle at the edge. From this, we conclude that if the internal DMI strength $D_{\text{A}}$ is large, then a strong canting away from the easy axis is to be expected close to the edge. If we add a non-zero boundary-induced DMI at the top and bottom surface ($D_{\text{S}}\neq 0$), then the magnetization will gain an additional twist in the $z$~direction close to the edge where the magnetization is canted away from the easy axis. For the sake of clarity, we assume that there is no boundary-induced DMI at the lateral surface ($D^{(1)}_{\text{S}}=0$). To relax the magnetization in the vicinity the edge of a film we use again the steepest gradient method. The obtained equilibrium magnetization at the edge is shown in Fig.~\ref{fig:crosssections}(b). Note how the relaxed magnetization twists along the $x$ as well as the $z$-direction close to the edge. When we compare the surface twist which occurs near the edge of a sample with the surface twists in a laterally infinite film induced by an in-plane applied field [shown in Fig.~\ref{fig:crosssections}(a)], we see that the magnitudes of the surface twists are comparable where the bulk magnetization has the same canting angle.

\section{Isolated domain wall\label{sec:wall}}

In this section, we discuss the surface twist in a straight isolated domain wall in a laterally extended film. As done previously, we consider an infinite film with thickness $d$. This time however, we assume that the magnetization~$m(x,z)$ is constant along the $y$ direction, but is allowed to vary in the $x$ and $z$ directions. Furthermore, we consider a nonzero internal DMI strength $|D_{\text{A}}|>0$, which causes the domain wall to be of the N\'eel type with a chirality fixed by the sign of $D_{\text{A}}$. Because we only concern ourselves with statics, the actual magnitude of the internal DMI has no influence on the profile of the domain wall. Under these assumptions, the system under study is two-dimensional and has only two free parameters: the symmetric DMI parameter $D_{\text{S}}$ and the thickness of the film $d$. The magnetization can be described by $m=(\cos\psi,0,\sin\psi)$ with the out-of-plane angle $\psi(x,z)$ being a function of $x$ and $z$. To determine the equilibrium state, we numerically minimize the free energy while taking into account the symmetry of the problem and the above assumptions.

\begin{figure}
    \centering
    \includegraphics{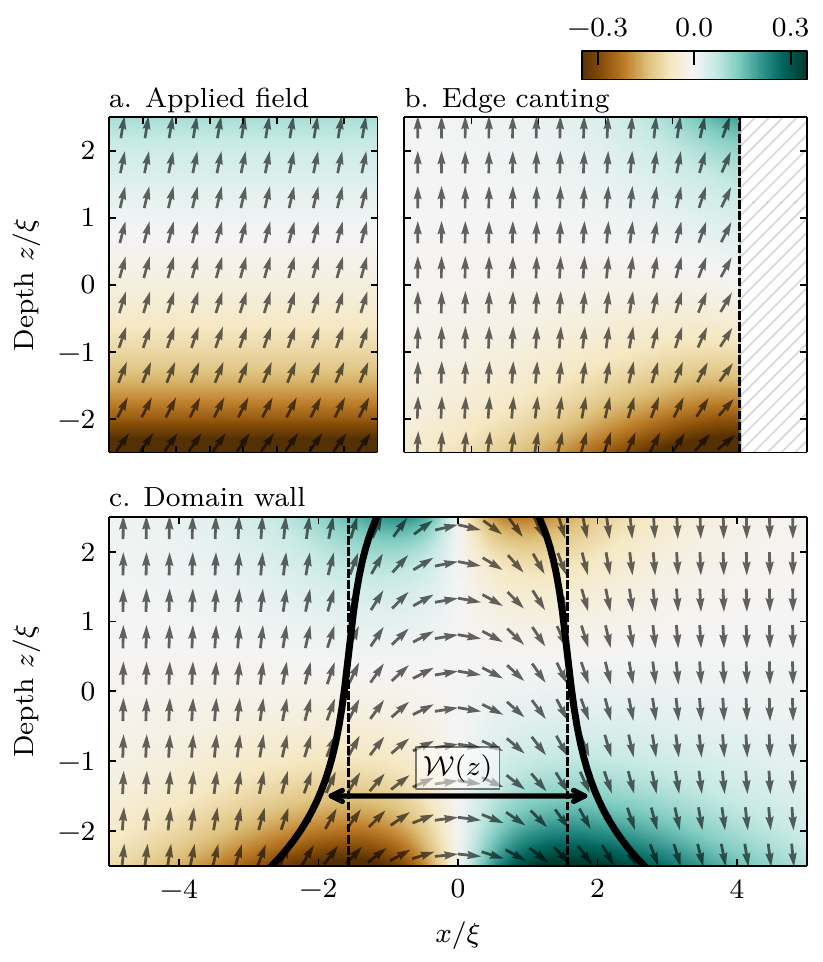} 
    \caption{Exemplary cross sections of the magnetization in a thin film, for the three different cases in which surface twists occur. (a) Quasi uniform state in a laterally infinite film with an applied field $\vec{H}^{\text{ext}}=0.5h_{\text{c}}\hat{e}_x$. (b) Canting at a sample edge without applied field. (c) Domain wall in a laterally infinite film without applied field. In all cases, the film has a thickness~$d=5\xi$, a symmetric DMI parameter~$D_{\text{S}}=0.6D_c$, and an asymmetric DMI parameter~$D_{\text{A}}=0.9D_c$. The color represents the angular difference $\Delta\psi$ with the magnetization in absence of the boundary-induced DMI ($D_{\text{S}}=0$). The distance between the indicated lines in (c) is the domain wall width~$\mathcal{W}(z)$ for $D_{\text{S}}=0.6D_c$ (solid line) and for $D_{\text{S}}=0$ (dotted line).}
    \label{fig:crosssections}
\end{figure}

\begin{figure}
    \centering
    \includegraphics{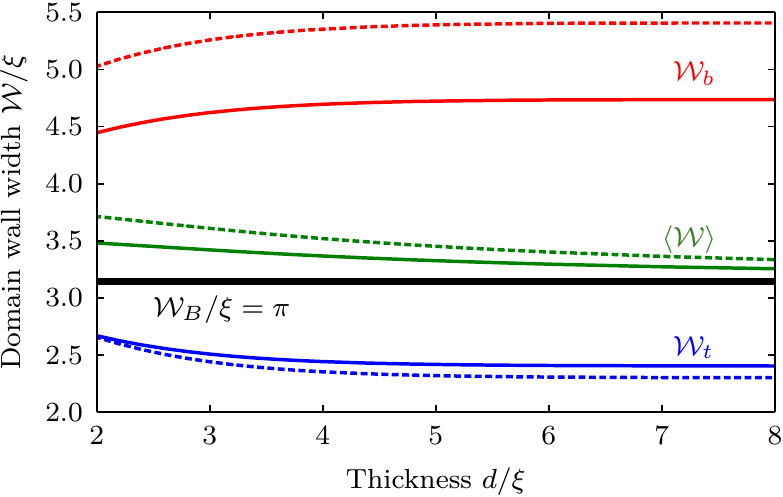} 
    \caption{The average domain wall width $\left<\mathcal{W}\right>$ and the domain wall width at the top $\mathcal{W}_t$ and bottom $\mathcal{W}_b$ of a thin film with symmetric DMI strength $D_{\text{S}}=0.5 D_{\text{c}}$ (solid line) and $D_{\text{S}}=0.6 D_{\text{c}}$ (dashed line), as a function of its thickness $d$.}
    \label{fig:wallwidth}
\end{figure}

Figure~\ref{fig:crosssections}(c) shows an example of a resulting cross section of a domain wall. As can be deduced from the BC given in Eq.~\eqref{eq:BC}, the surface twist is only present in regions where the magnetization is neither parallel, nor orthogonal to the interface. Thus, the surface twist occurs at the left and at the right of the domain wall, but not at its center. Due to the surface twist, the domain wall width varies along the $z$~direction. In order to quantify this dependence, we use the following definition of the domain wall width:
\begin{equation}
    \mathcal{W}(z)= \int \sqrt{1-m^2_z(x,z)} \,d x = \int \left|\cos\psi(x,z)\right| \,d x. \label{eq:wallwidth}
\end{equation}
For very thick films $(d\rightarrow\infty)$, one can assume that the shape of the domain wall in the bulk is not affected by the boundary-induced DMI. In that case, the domain wall profile\cite{Coey} is given by the Gudermannian function $\psi(x)=\gd(x) = 2 \arctan [\exp(x/\xi)] -\pi/2\ $. Using the definition given in Eq.~\eqref{eq:wallwidth}, we obtain the domain wall width in the bulk~$\mathcal{W}_B = \pi\xi$. If we add the boundary-induced DMI, the domain wall width will differ from $\pi\xi$ and will vary along the $z$~direction. The average domain wall width and the domain wall widths at the top and bottom surface are shown in Fig.~\ref{fig:wallwidth} as a function of the film thickness~$d$ and the boundary-induced DMI strength.

In case of a positive boundary-induced DMI~$D_{\text{S}} > 0$, N\'eel domain walls are wider at the bottom and thinner at the top surface, independently of the chirality of the domain wall. The opposite is true if the boundary-induced DMI is negative. For increasing film thicknesses, the difference between the domain wall width at the top and bottom surface increases until the domain wall width eventually converges to a fixed width which is proportional to the DMI strength~$D_{\text{S}}$. The broadening of the domain wall at the bottom surface has a stronger extent than the narrowing at the top surface. Consequently, the average domain wall width in films of finite thickness will be larger than the domain wall width in absence of a boundary-induced DMI: $\left<\mathcal{W}\right> > \mathcal{W}_B$. This effect is more pronounced in thin films ($d<5\xi$).

\section{Isolated Skyrmion\label{sec:skyrmion}}

In this section, we analyze the influence of the boundary-induced DMI on the profile of an isolated skyrmion. To this end, we suppose that the magnetization has a circular symmetry and has a purely N\'eel character. This is a reasonable assumption because in $C_{\infty v}$ systems a Bloch-like twist leads to an increase of the free magnetic energy. Using this assumption, the magnetization can be described by
\begin{equation}
    \vec{m} = (\cos\phi\cos\psi(r,z),\sin\phi\cos\psi(r,z),\sin\psi(r,z))
\end{equation}
in the polar coordinate system $(r,\phi,z)$. Due to the circular symmetry constraint, the magnetization is fully determined by the angle $\psi(r,z)$. The total energy of the system is given by
\begin{equation}
    E[\psi] = 2\pi\int^{R}_{0} \int_0^d \varepsilon(r,z) r\, \,dr\,dz,
\end{equation}
with the magnetic free energy density
\begin{equation}
    \begin{aligned}
        \varepsilon ={} & A \left((\partial_r \psi)^2 + (\partial_z \psi)^2 + \frac{\cos^2\psi}{r^2}\right)
            + K_{\text{e}} \cos^2\psi \\
        + & D_A \left( \partial_r\psi - \frac{\cos\psi\sin\psi}{r} \right)
            -  2 D_{\text{S}} \sin\psi\cos\psi \partial_z \psi
    \end{aligned}
\end{equation}
Note that, also here, the term with the symmetric DMI parameter $D_{\text{S}}$ can be integrated over the $z$ direction analytically, which reduces this term to a top and a bottom surface energy term. Finally, we minimize the energy functional $E[\psi]$ numerically.

\begin{figure}
    \centering
    \includegraphics{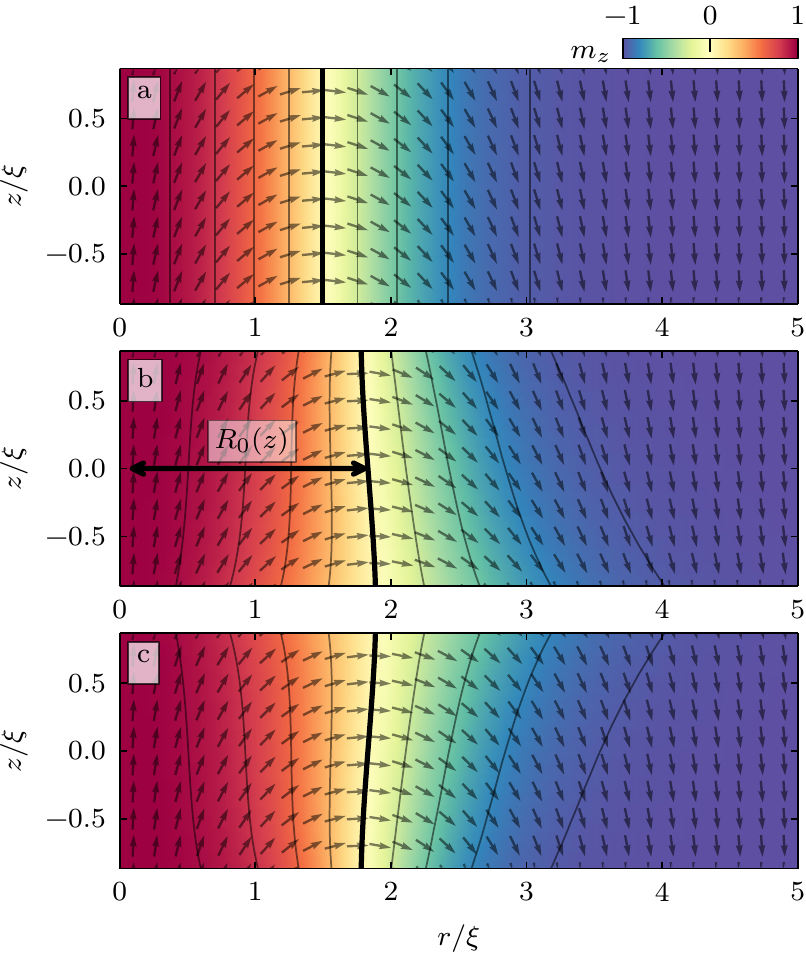} 
    \caption{Skyrmion profile in a film with DMI strength $D_{\text{A}}=0.8D_c$ and thickness $d=2\xi$ (a) without boundary-induced DMI, and (b) with boundary-induced DMI $D_{\text{S}}=0.5D_{\text{c}}$, and (c) $D_{\text{S}}=-0.5D_{\text{c}}$. A 3D representation of the deformed skyrmion is readily shown in Fig.~\ref{fig:demo}}
    \label{fig:skyrmioncrosssection}
\end{figure}

\begin{figure}
    \centering
    \includegraphics{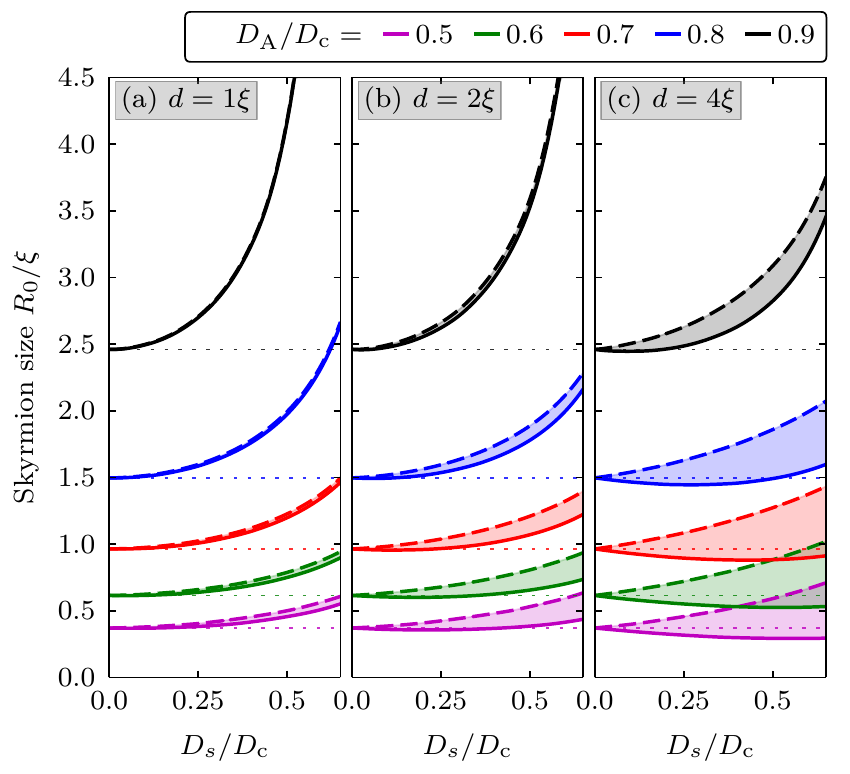} 
    \caption{The radius $R_0(z)$ of an isolated skyrmion in a film with thickness $d$ increases monotonically from the top (solid line) to the bottom surface (dashed line). The skyrmion size at the top and the bottom surface are shown as function of the boundary-induced DMI $D_{\text{S}}$ (horizontal axis) and the bulk DMI $D_A$ (color). The dotted reference line is the skyrmion size in absence of boundary-induced DMI ($D_{\text{S}}=0$).}
    \label{fig:skyrmionsize}
\end{figure}

Figure~\ref{fig:skyrmioncrosssection} shows a typical profile of a relaxed skyrmion. As we assume a radial profile, we only plot the magnetization texture of the skyrmion from its center to its boundary, i.e.\ half of the front of Fig.~\ref{fig:demo}. Similar to the straight isolated domain wall, we observe narrowing/broadening of the domain wall at the top/bottom surface due to the boundary-induced DMI. The asymmetry of the deformation at the top and bottom surface causes an increase in the skyrmion size and a 3D deformation of the skyrmion profile. We performed a parameter study in order to check how these deviations depend on the thickness of the film and on the DMI strengths $D_{\text{S}}$ and $D_{\text{A}}$. The results of this study are shown in Fig.~\ref{fig:skyrmionsize}.

The skyrmion size in very thin films is virtually constant along the $z$~direction [see Fig.~\ref{fig:skyrmionsize}(a)], whereas it varies strongly along the $z$~direction in thick films [see Fig.~\ref{fig:skyrmionsize}(a)], which leads to a skewed skyrmion domain wall which is clearly visible in Fig.~\ref{fig:skyrmioncrosssection}. More interestingly, the average skyrmion size depends strongly on both the inner DMI strength~$D_{\text{A}}$ and the boundary-induced DMI strength~$D_{\text{S}}$. It is already known that the skyrmion size increases for an increasing internal DMI strength~$D_{\text{A}} < D_{\text{c}}$, and that for an internal DMI strength larger than the critical value~$D_{\text{c}}$, a skyrmion in an extended film will expand and/or deform in order to maximize its circumference. Due to this unwieldy behavior for a strong internal DMI, we limited our parameter study to internal DMI strengths below the critical value ($D_{\text{A}}<D_{\text{c}}$). For a given internal DMI strength~$D_{\text{A}}$, the average skyrmion size increases for an increasing boundary-induced DMI strength~$D_{\text{S}}$. This dependence is especially notable for internal DMI strengths just below the critical value [see black lines in Fig.~\ref{fig:skyrmionsize}]. If we look closely at Fig.~\ref{fig:skyrmionsize}(a) and Fig.~\ref{fig:skyrmionsize}(b), the results seem to suggest that in thin films the skyrmion size diverges when the internal DMI as well as the boundary-induced DMI are strong, but still lower than the critical DMI strength~$D_{\text{c}}$.

One may expect that the boundary-induced DMI has an influence on the stability of a skyrmion. Atomistic simulations of skyrmions have shown that a skyrmion can collapse into the ferromagnetic ground state \cite{Bessarab2015,Rohart2016,Lobanov2016,Siemens2016,Stosic2017,Cortes-Ortuno2017}. In the case of a weak internal DMI or a perpendicular applied field in the opposite direction as the center magnetization of the skyrmion, the skyrmion is small and prone to collapse into the ferromagnetic ground state. Vice versa, for an increasing internal DMI strength or an increasing perpendicular applied field in the same direction as the center magnetization, the skyrmion size increases together with the skyrmion stability. In the results of our parameter study of the 3D deformed skyrmions, we see that the skyrmion size increases with both the internal DMI as well as the boundary-induced DMI. This suggests that not only the internal DMI or a perpendicular applied field, but also the boundary-induced DMI, counteracts the collapse of skyrmions in thin films.

Finally, one could also expect that the boundary-induced DMI has an effect on skyrmion dynamics. The change in the skyrmion and domain wall profile will most likely influence the eigenmodes and the motion under an applied spin-polarized current, which might be of importance for the development of e.g.\ racetrack memory \cite{Fert2013,tomasello2014,Zhang2015,Iwasaki2014}.

\section{Conclusions\label{sec:conclusions}}

To summarize, in this paper we have analyzed the influence of the recently predicted boundary-induced Dzyaloshinskii-Moriya interaction (DMI) on the magnetization in a uniform state, a domain wall or a skyrmion, or in vicinity of an edge, in ferromagnetic films with a $C_{\infty v}$ symmetry and a perpendicular magnetic anisotropy. We have rendered pronounced effects that lead to novel and peculiar profile changes along the thickness of the film for all considered magnetic textures. Among the most notable effects is the deformation of the domain wall between the surfaces of the film, as well as the increase of the average skyrmion size, which suggests that the boundary-induced DMI can positively contribute to the skyrmion stability.

In this paper, we solely focused on the effects of the boundary-induced DMI at the top and bottom surface of thin films with a $C_{\infty v}$ symmetry. In such systems, the internal DMI favors rotation of the magnetization in the $x$ and $y$~direction but does not induce a change in the $z$~direction. Hence, all variations of the magnetization in the $z$~direction can be attributed to the presence of the boundary-induced DMI, which makes these systems an ideal first study case. However, one can expect to find similar phenomena owing to boundary-induced DMI in other systems as well. For instance, it is worth to investigate the case in which the internal DMI of a film already leads to a variation in the $z$-direction and where the boundary-induced DMI results in an additional twist. Also in heterochiral magnets, a boundary-induced DMI can occur at interfaces where the internal DMI strength changes\cite{mulkers2017,Stosic2017}, which could lead to an additional twist to the already predicted spin canting at the interface.

One could also expect that the variation of the magnetization along the film's normal can lead to non-trivial excitations in this direction and a peculiar dynamics when applying a perpendicular spin-polarized current. Studying the magnetization dynamics would not only be interesting for the design of spintronic devices, but might also provide an indirect way to measure the boundary-induced DMI strength.


\begin{acknowledgements}
The authors thank Matthias Sitte and Andr\'e Thiaville for fruitful discussions. This work was supported by the Fonds Wetenschappelijk Onderzoek (FWO-Vlaanderen) through Project No. G098917N and the German Research Foundation (DFG) under the Project No. EV 196/2-1. J. L. is supported by the Ghent University Special Research Fund with a BOF postdoctoral fellowship.
\end{acknowledgements}

\appendix

\section{Demagnetization of quasi-uniform magnetized film}\label{sec:demaguniform}

In this appendix we prove that the demagnetization of an infinite film can be exactly described by a shape anisotropy if one assumes that the magnetization~$\vec{M}(z)$ varies only along the $z$ direction.

The demagnetization field is given by
\begin{equation}
    \vec{H}^{\text{demag}} = \vec{H}^{\Omega} + \vec{H}^{\partial\Omega},
\end{equation}
with a contribution from volume charges:
\begin{equation}
    \vec{H}^{\Omega}(\vec{r}) = - \frac{1}{4\pi} \int_{\Omega} \nabla\cdot \vec{M}(\vec{r}') \frac{\vec{r}-\vec{r}'}{\left|\vec{r}-\vec{r}'\right|^3} \,d^3 \vec{r}',
\end{equation}
and a contribution from the surface charges:
\begin{equation}
    \vec{H}^{\partial\Omega}(\vec{r}) = \frac{1}{4\pi}\int_{\delta\Omega} \vec{n}(\vec{r}')\cdot \vec{M}(\vec{r}')\frac{\vec{r}-\vec{r}'}{\left|\vec{r}-\vec{r}'\right|^3} \,d^2 \vec{r}'.
\end{equation}
When considering a magnetization which varies only along the $z$ direction, then the integration over $x$ and $y$ in the volume and surface integrals can be carried out as follows
\begin{equation}
    \int_{-\infty}^{+\infty} \int_{-\infty}^{+\infty}
    \frac{\vec{r}-\vec{r}'}{\left|\vec{r}-\vec{r}'\right|^3}
    \,dx\,dy
    = 2\pi \frac{z-z'}{\left|z-z'\right|} \hat{e}_z.
\end{equation}
For the volume integral we still need to integrate over the depth of the film. This yields
\begin{equation}
    \vec{H}^{\Omega}(z) = \left[\frac{M^t_z + M^b_z}{2} - M_z(z) \right] \hat{e}_z,
\end{equation}
where the upper indices $t$ and $b$ denote the magnetization at the top and bottom of the film respectively. The surface integral reduces to
\begin{equation}
    \vec{H}^{\partial\Omega}(z) =  \frac{M^t_z + M^b_z}{2}\hat{e}_z.
\end{equation}
Finally, we obtain the demagnetization energy density
\begin{equation}
    \varepsilon^{\text{demag}} = -\frac{\mu_0}{2} \vec{M}\cdot \vec{H}^{\text{demag}} = \frac{\mu_0 M^2_{\text{s}}}{2} m^2_z,
\end{equation}
which is equivalent to the energy density of a film with uniaxial anisotropy with an hard axis perpendicular to the film and anisotropy constant $-\mu_0 M^2_{\text{s}}/2$.

\bibliography{main.auto,coey}

\end{document}